\documentclass[fleqn,12pt,twoside]{article}

\usepackage{graphicx}
\usepackage[headings]{espcrc1}
\usepackage{bm}
\usepackage{epsfig}
\usepackage{amsmath}
\usepackage{amssymb}

\title{Kolmogorov-Sinai entropy from recurrence times} 

\author{M.  S.  Baptista\address[aberdeen]{Institute for Complex
Systems and Mathematical Biology, King's College, University of
Aberdeen, AB24 3UE Aberdeen, United Kingdom}\address[CMUP]
{Centro de Matem\'atica da Universidade do Porto, Rua do Campo
Alegre 687, 4169-007 Porto, Portugal}\thanks{partially supported  
by the ``Funda\c c\~ao para a Ci\^encia e Tecnologia'' (FCT).}, 
E. J. Ngamga\address[potsdam]{Potsdam Institute
for Climate Impact Research, Telegraphenberg, 14412 Potsdam, Germany}
\thanks{partially supported by SFB555.},
Paulo R. F. Pinto\addressmark[CMUP]\thanks{partially supported  
by the ``Funda\c c\~ao para a Ci\^encia e Tecnologia'' (FCT).}, 
 Margarida Brito\addressmark[CMUP],
 J. Kurths\addressmark[aberdeen,potsdam]\thanks{partially supported by SFB555.}}

\runtitle{Kolmogorov-Sinai entropy from recurrence times}
\runauthor{M. S. Baptista}
\begin{document}

\maketitle 

\begin{abstract}
{\bf ABSTRACT:} Observing how long a dynamical system takes to return to
  some state is one of the most simple ways to model and quantify its
  dynamics from data series.  This work proposes two formulas to
  estimate the KS entropy and a lower bound of it, a sort
  of Shannon's entropy per unit of time, from the recurrence times of
  chaotic systems. One formula provides the KS entropy and is more
  theoretically oriented since one has to measure also the low
  probable very long returns. The other provides a lower bound for the
  KS entropy and is more experimentally oriented since one has to
  measure only the high probable short returns. These formulas are a
  consequence of the fact that the series of returns do contain the
  same information of the trajectory that generated it. That suggests
  that recurrence times might be valuable when making models of
  complex systems.
\end{abstract}

\vspace{1.0cm}

\section{Introduction}

Recurrence times measure the time interval a system takes to return to
a neighborhood of some state, being that it was previously in some
other state. Among the many ways time recurrences can be defined, two
approaches that have recently attracted much attention are the first
Poincar\'e recurrence times (FPRs) \cite{poincare} and the recurrence
plots (RPs)
\cite{ruelle}.
 
While Poincar\'e recurrences refer to the sequence of time intervals
between two successive visits of a trajectory (or a signal) to one
particular interval (or a volume if the trajectory is high
dimensional), a recurrence plot refers to a visualization of the
values of a square array which indicates how much time it takes for
two points in a trajectory with $M$ points to become neighbors
again. Both techniques provide similar results but are more
appropriately applicable in different contexts. While the FPRs are
more appropriated to obtain exact dynamical quantities (Lyapunov
exponents, dimensions, and the correlation function) of dynamical
systems
\cite{FPR}, the RPs are more oriented to
estimate relevant quantities and statistical characteristics of data
coming from complex systems \cite{RP}.

The main argument in order to use recurrence times to model complex
systems \cite{baptista_DNA} is that one can easily have experimental
access to them. In order to know if a model can be constructed from
the recurrence times, it is essential that at least the series of
return times contains the same amount of information generated by the
complex system, information being quantified by the entropy.

Entropy is an old thermodynamic concept and refers to the disorganized
energy that cannot be converted into work. It was first mathematically
quantified by Boltzmann in 1877 as the logarithm of the number of
microstates that a gas occupies. More recently, Shannon
\cite{shannon} proposed a more general way to measure entropy $H_S$
in terms of the probabilities $\rho_i$ of all possible $i$ states of a
system:{\small
\begin{equation}
H_S= -\sum_i \rho_i \log{(\rho_i)}.
\label{se}
\end{equation}}

Applied to non-periodic continuous trajectories, e.g. chaotic
trajectories, $H_S$ is an infinite quantity due to the infinitely many
states obtained by partitioning the phase space in arbitrarily small
sites. Therefore, for such cases it is only meaningful to measure
entropy relative to another trajectory. In addition, once a dynamical
system evolves with time, it is always useful for comparison reasons
to measure its entropy production per unit of time.

Such an ideal entropy definition for a dynamical system was introduced
by Kolmogorov in 1958 \cite{kolmogorov} and reformulated by Sinai in
1959. It is known as the Kolmogorov-Sinai (KS) entropy, denoted by
$H_{KS}$, basically the Shannon's entropy of the set per unit of time
\cite{KS-entropy}, and it is the most successful invariant quantity
that characterize a dynamical system \cite{peter}. However, the
calculation of the KS entropy to systems that might possess an
infinite number of states is a difficult task, if not impossible. For
a smooth chaotic system \cite{young} (typically happens for
dissipative systems that present an attractor), Pesin \cite{pesin}
proved an equality between $H_{KS}$ and the sum of all the positive
Lyapunov exponents. However, Lyapunov exponents are difficult or even
impossible to be calculated in systems whose equations of motion are
unknown. Therefore, when treating data coming from complex systems,
one should use alternative ways to calculate the KS entropy, instead
of applying Pesin's equality.
                                     
Methods to estimate the correlation entropy, $K_2$, a lower bound of
$H_{KS}$, and to calculate $H_{KS}$ from time series were proposed in
Refs. \cite{grassberger,cohen}. In Ref. \cite{grassberger} $K_2$ is
estimated from the correlation decay and in Ref. \cite{cohen} by the
determination of a generating partition of phase space that preserves
the value of the entropy. But while the method in Ref.
\cite{grassberger} unavoidably suffers from the same difficulties
found in the proper calculation of the fractal dimensions from data
sets, the method in Ref. \cite{cohen} requires the knowledge of the
generating partitions, information that is not trivial to be extracted
from complex data \cite{baptista_PRL2006}. In addition, these two
methods and similar others as the one in Ref. \cite{wolf} require the
knowledge of a trajectory. Our work is devoted to systems whose
trajectory cannot be measured.

A convenient way of determining all the relevant states of a system
and their probabilities (independently whether such a system is
chaotic) is provided by the FPRs and the RPs. In particular to the
Shannon's entropy, in Refs. \cite{Trulla,Faure,letellier,RP} ways were
suggested to estimate it from the RPs. In Refs. \cite{Trulla,Faure,RP}
a subset of all the possible probabilities of states, the
probabilities related to the level of coherence/correlation of the
system, were considered in Eq. (\ref{se}).  Therefore, as pointed out
in Ref. \cite{letellier}, the obtained entropic quantity does not
quantify the level of disorganization of the system. Remind that
unavoidably Shannon's entropy calculated from RPs or FPRs depends on
the resolution with which the returns are measured.

The main result of this contribution is to show how to easily estimate
the KS-entropy from return times, without the knowledge of a
trajectory. We depart from similar ideas as in Refs.
\cite{Trulla,Faure,letellier,RP} and show that the KS entropy is the
Shannon entropy [in Eq. (\ref{se})] calculated considering the
probabilities of all the return times observed divided by the length
of the shortest return measured. This result is corroborated with
simulations on the logistic map, the H\'enon map, and coupled maps. We
also show how to estimate a lower bound for the KS entropy using for
that the returns with the shortest lengths (the most probable
returns), an approach oriented to the use of our ideas in experimental
data. Finally, we discuss in more details the intuitive idea of
Lettelier \cite{letellier} to calculate the Shannon's entropy from a RP
and show the relation between Letellier's result and the KS
entropy.


\section{Estimating the KS entropy from time returns}

Let us start with some definitions. By measuring two subsequent
returns to a region, one obtains a series of time intervals (FPRs)
denoted by $\tau_i$ (with $i=1,\ldots,N$). The characterization of the
FPRs is done by the probability distribution
$\rho(\tau,{\mathcal{B}})$ of $\tau_i$, where ${\mathcal{B}}$
represents the volume within which the FPRs are observed. In this
work, ${\mathcal{B}}$ is a $D$-dimensional box, with sides
$\epsilon_1$, and $D$ is the phase space dimension of the system being
considered. We denote the shortest return to the region
${\mathcal{B}}$ as $\tau_{min}({\mathcal{B}})$.

Given a trajectory $\{\bf x_i\}_{i=1}^M$, the recurrence plot is a
two-dimensional graph that helps the visualization of a square array
$R_{ij}$:{\small
\begin{equation}
R_{ij}=\theta(\epsilon_2 - \| {\bf x_i} - {\bf x_j} \|)
\label{recurrent_plot}
\end{equation}}
\noindent
where $\epsilon_2$ is a predefined threshold and $\theta$ is the
Heaviside function \cite{ruelle}. In the coordinate $(i,j)$ of the RP
one plots a black point if $R_{ij}=1$, and a white point otherwise.

There are many interesting ways to characterize a RP, all of them
related to the lengths (and their probabilities of occurrence) of the
diagonal, horizontal, and vertical segments of {\it recurrent} points
(black points) and of {\it nonrecurrent} points (white
points). Differently from Ref.
\cite{letellier} where it was used the nonrecurrent diagonal segments, we consider 
here the vertical nonrecurrent and recurrent segments because they
provide a direct link to the FPRs \cite{coment_equivalencia}.

Given a column $i$, a vertical segment of $Q$ white points starting at
$j=p$ and ending at $j=p+Q-1$, indicates that a trajectory previously
in the neighborhood of the point ${\bf x}_i$ returns to it firstly
after $Q+1$ iterations in the neighborhood of the point ${\bf x}_i$,
basically the same definition as the FPR to a volume centered at ${\bf
x}_i$. However, the white points represent returns to the neighborhood
of ${\bf x}_i$ which are larger than 1. In order to obtain the returns
of length 1, one needs to use the recurrent segments, the segments
formed by black points. A recurrent vertical segment at the column
$i$, starting at $j=p$ and ending at $j=p+Q$, means that it occurred
$Q$ first returns of length 1 to the neighborhood of the point ${\bf
x}_i$. The probability density of the return times observed in the RP
is represented also by $\rho(\tau,{\mathcal{B}})$. It is constructed
considering the first returns observed in all columns of the RP and it
satisfies $\int \rho(\tau,{\mathcal{B}}) d\tau =1$.

Notice that the Shannon's entropy of first returns of non-periodic
continuous systems becomes infinite
\cite{shannon_entropy_FPR} as the size $\epsilon$ of the volume 
${\mathcal{B}}$ approaches zero. For chaotic systems (as well as for
stochastic systems) the reason lies on the fact that the probability
density $\rho(\tau,{\mathcal{B}})$ approaches the exponential form
$\mu e^{-\mu \tau}$ \cite{exponential}, where $\mu$ is the probability
of finding the trajectory within the volume ${\mathcal{B}}$.

Placing in Eq. (\ref{se}) the probabilities of returns
$\rho(\tau,{\mathcal{B}})$, we can write that $H_{KS} = H_S/T$, where
$T$ is some characteristic time of the returns \cite{KS-entropy} that
depends on how the returns are measured.
For the FPRs there exists three characteristic times: the shortest,
the longest and the average return.  The quantity $T$ cannot be the
longest return since it is infinite. It cannot be the average return, 
since one would arrive to $H_{KS}
\cong \mu \log{(\mu)}$ which equals zero as $\epsilon \rightarrow 0$. 
Therefore, $T=\tau_{min}$ is the only remaining reasonable
characteristic time to be used which lead us to
\begin{equation}
H_{KS}({\mathcal{B}}[\epsilon])=\frac{1}{\tau_{min}({\mathcal{B}}[\epsilon])}
\sum_{\tau} \rho(\tau,{\mathcal{B}}[\epsilon])
\log{\left(\frac{1} {\rho(\tau,{\mathcal{B}}[\epsilon])} \right)}.
\label{HKS_prob}
\end{equation}

For uniformly hyperbolic chaotic systems (tent map, for example), we
can prove the validity of Eq.  (\ref{HKS_prob}). From Ref.
\cite{paulo} we have that
\begin{equation}
H_{KS}=-\lim_{\epsilon \rightarrow 0}\frac{1}{\tau_{min}} \log(\rho(\tau_{min},{\mathcal{B}}[\epsilon]))
\end{equation}
a result derived from the fact that the KS entropy exponentially
increases with the number of unstable periodic orbits embedded in the
chaotic attractor. Since $\rho(\tau,\epsilon) \rightarrow \mu e^{-\mu
  \tau}$ as $\epsilon \rightarrow 0$, assuming $\tau_{min}$ to be very
large, and noticing that $\int -\mu e^{-\mu \tau} log{[\mu e^{-\mu
    \tau}]} d\tau= -log{[\mu]}+1$, assuming that $\tau_{min}
\rightarrow \infty$ and noticing that for such systems
$\mu[{\mathcal{B}}]=\rho(\tau_{min},\epsilon)$, we finally arrive that
\begin{equation}
- \frac{1}{\tau_{min}} \log{[\rho(\tau_{min})]} = 
- \frac{1}{\tau_{min}}
\sum_{\tau} \rho(\tau)
\log{[\rho(\tau)]}
\label{demonstration}
\end{equation}
\noindent
and therefore, the right-hand side of Eq. (\ref{HKS_prob}) indeed
reflects the KS entropy. But notice that Eq. (\ref{HKS_prob}) is being
applied not only to non-uniformly hyperbolic systems (Logistic and
H\'enon maps) but also to higher dimensional systems (two coupled
maps).

This result can also be derived from Ref. \cite{benoit} where it was
shown that the positive Lyapunov exponent $\lambda$ in hyperbolic 1D
maps is
\begin{equation}
\lambda = \lim_{\epsilon
\rightarrow 0} \frac{-log{[\mu(\epsilon)]}}{\tau_{min}(\mathcal{B}[\epsilon])}. 
\label{benoit}
\end{equation}
Since $\rho(\tau,\epsilon)
\rightarrow \mu e^{-\mu \tau}$ as 
$\epsilon \rightarrow 0$, and using that $\lambda=H_{KS}$ (Pesin's
equality), and finally noticing that $\int -\mu e^{-\mu \tau} log{[\mu
e^{-\mu
\tau}]} d\tau= -log{[\mu]}+1$, one can arrive to the conclusion that 
$T=\tau_{min}$ in Eq. (\ref{HKS_prob}).

The quantity in Eq. (\ref{HKS_prob}) is a local estimation of the KS
entropy. To make a global estimation we can define the average
\begin{equation}
\langle H_{KS} \rangle = \frac{1}{L} \sum_{\mathcal{B}(\epsilon)} H_{KS} [\mathcal{B}(\epsilon)]
\label{average_HKS}
\end{equation}
\noindent
representing an average of $H_{KS} [\mathcal{B}(\epsilon)]$ calculated
considering $L$ different regions in phase space.

In order to estimate the KS entropy in terms of the probabilities
obtained from the RPs, one should use $T=\langle \tau_{min} \rangle$,
i.e., replace $\tau_{min}$ in Eq. (\ref{HKS_prob}) by $\langle
\tau_{min} \rangle$, where $\langle
\tau_{min} \rangle =\frac{1}{M} \sum_i
\tau_{min}(i)$, the average value of the shortest return 
observed in every column of the RP. The reason to work with an average
value instead of using the shortest return considering all columns of
the RP is that every vertical column in the RP defines a shortest
return $\tau_{min}(i)$ ($i=1,\ldots,M$), and it is to expect that
there is a nontypical point $i$ for which $\tau_{min}(i)=1$. 

Imagining that the RP is constructed considering arbitrarily small
regions ($\epsilon_2 \rightarrow 0$) and that we could treat an
arbitrarily long data set, the column of the RP which would produce
$\tau_{min}=1$ would be just one out of infinite others which produce
$\tau_{min} >> 1$. There would be also a finite number of columns
which would produce $\tau_{min}$ of the order of one (but larger than
one), but also those could be neglect when estimating the KS-entropy
from the RPs. The point we want to make in here is that the possible
existence of many columns for which one has $\tau_{min}=1$ are a
consequence of the finite resolution with which one constructs a RP.
In order to minimize such effect in our calculation we just ignore the
fact that we have indeed found in the RP $\tau_{min}=1$, and we
consider as $\tau_{min}$ any return time longer than 1 as the minimal
return time. In fact, neglecting the existence of returns of length
one is a major point in the work of Ref. \cite{letellier}, since there
only the nonrecurrent diagonal segments are considered
\cite{coment_equivalencia}, and thus, the probability of having a
point returning to its neighborhood after one iteration is zero.

From the conditional probabilities of returns, a lower bound for the
KS entropy can be estimated in terms of the FPRs and RPs by
\begin{equation}
H_{KS}({\mathcal{B}}[\epsilon]) \geq - \frac{1}{n} \sum_{i=1}^n
\frac{1}{P_i} \frac{\rho(\tau_i+P_i)}{\rho(\tau_i)}
\log{\left[\frac{\rho(\tau_i+P_i)}{\rho(\tau_i)}\right]}
\label{HKS_cond}
\end{equation}
\noindent
where we consider only the returns $\tau_i$ for which
$\rho(\tau_i+P_i)/\rho(\tau_i)>0$ and $\tau_i+P_i < 2\tau_{min}$, with
$P_i \in \mathcal{N}$.

The derivation of Eq. (\ref{HKS_cond}) is not trivial because it
requires the use of a series of concepts and quantities from the
Ergodic Theory. In the following, we describe the main steps to arrive
at this inequality.

First we need to understand the way the KS-entropy is calculated via a
spatial integration. In short, the KS-entropy is calculated using the
Shannon's entropy of the conditional probabilities of trajectories
within the partitions of the phase space as one iterates the chaotic
system backward
\cite{ruelle}. More rigorously, denote a phase space partition
$\delta_N$. By a partition we refer to a space volume but that is
defined in terms of Markov partitions. Denote $S$ as $S=S_0 \cap S_1
\cap S_{k-1}$ where $S_j \in F^{-j}
\delta_N$ ($j=0,\ldots,k-1$), where $F$ is a chaotic transformation. 
Define $h_{N}(k) = \frac{\mu(S \cap S_k)}{\mu(S)} log{\frac{\mu(S
\cap S_k)}{\mu(S)}}$ and $\mu(S)$ represents the probability measure 
of the set $S$. The KS-entropy is defined as $H_{KS} =
\lim_{l \rightarrow \infty} \frac{1}{l} \sum_{k=0}^{l-1}
\int
\rho(dx) h_{N}(k)$, where the summation is taken over 
$l$ iterations.

Assume now that the region ${\mathcal{B}}$ represents the good
partition $\delta_N$.  The region $S_j$ is the result of
$F^{-j}\delta_N$, i.e., a $j$-th backward iteration of
${\mathcal{B}}$. So, clearly, if one applies $j$ forward iterations to
$S_j$, then $F^{j}S_j \rightarrow {\mathcal{B}}$. The quantities
$\mu(S \cap S_k)$ and $\mu(S)$ refer to the measure of the chaotic
attractor inside $S
\cap S_k$ and $S$, respectivelly. By measure we mean the 
natural measure, i.e. the frequency with which a typical trajectory
visits a region. ${\mu(S \cap S_k)}$ refers to the measure that
remained in ${\mathcal{B}}$ after $k$ iterations and ${\mu(S)}$ the
measure that remained in ${\mathcal{B}}$ after $k-1$ iterations.

\begin{figure}
\hbox{\psfig{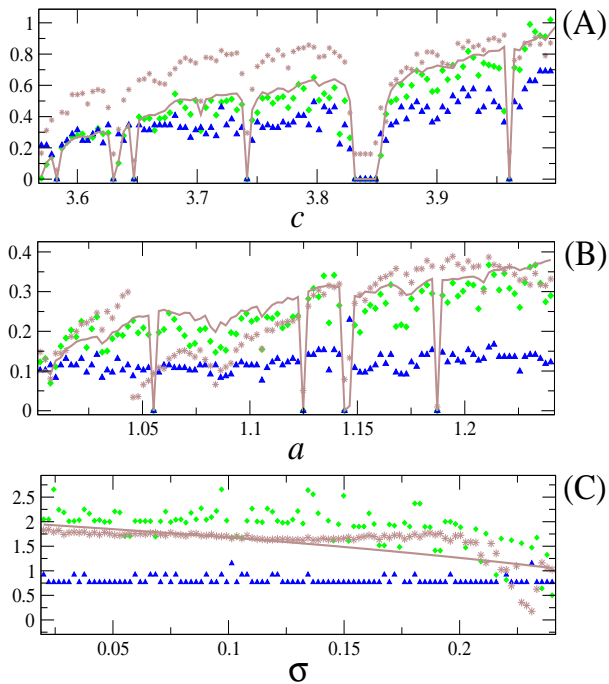}}
\caption{[color online] Results from Eq. (\ref{HKS_prob}) and 
(\ref{benoit}). The probability function $\rho(\tau,{\mathcal{B}})$ of
the FPRs (RPs) were obtained from a series of 500.000 FPRs (from a
trajectory of length 15.000 points).  The brown line represents the
values of the positive Lyapunov exponent. In (A) we show results for
the Logistic map as we vary the parameter $c$, $\epsilon_2=0.002$ for
the brown stars and $\epsilon_1$=0.001 for the green diamonds. In (B)
we show results for the H\'enon map as we vary the parameter $a$ for
$b$=0.3, $\epsilon_2=[0.002-0.03]$ for the brown stars and
$\epsilon_1$=0.002 for all the other results, and in (C) results for
the coupled maps as we vary the coupling strength $\sigma$,
$\epsilon_2$=0.05 for the brown stars and $\epsilon_1$=0.02 for green
diamonds.}
\label{fig1}
\end{figure}

For $k
\rightarrow \infty$, we have that $\frac{\mu(S
\cap S_k)}{\mu({\mathcal{B}})} \rightarrow \mu({\mathcal{B}})$.  Also 
for finite values of $k$, one has that 
$\frac{\mu(S \cap S_k)}{\mu({\mathcal{B}})} \approx \mu({\mathcal{B}})$. 
For any finite $k$, 
we can split this fraction into two components: $\frac{\mu(S
\cap S_k)}{\mu({\mathcal{B}})} = \mu_{REC}(k,{\mathcal{B}}) + \mu_{NR}(k,{\mathcal{B}})$. 
$\mu_{REC}$ refers to the measure in ${\mathcal{B}}$ associated with
unstable periodic orbits (UPOs) that return to ${\mathcal{B}}$, after
$k$ iteration of $F$, at least twice or more times. $\mu_{NR}$ refers
to the measure in ${\mathcal{B}}$ associated with UPOs that return to
${\mathcal{B}}$ only once.

As it is shown in Ref. \cite{paulo}, $\rho(\tau,{\mathcal{B}}) =
\mu_{NR}(\tau,{\mathcal{B}}) $, which in other words means
that the probability density of the FPRs in ${\mathcal{B}}$ is given
by $\mu_{NR}(k,{\mathcal{B}})$. But, notice that for $\tau <
2\tau_{min}$, $\mu_{REC}(k,{\mathcal{B}})=0$ since only returns
associated with UPOs that return once 
can be observed inside ${\mathcal{B}}$, and therefore
$\rho(\tau,{\mathcal{B}})$ = $\frac{\mu(S
\cap S_{\tau})}{\mu({\mathcal{B}})}$, if $\tau < 2\tau_{min}$. 
Consequently, we have that $\frac{\mu(S \cap S_{\tau})}{\mu(S)}=
\frac{\rho(\tau,{\mathcal{B}})}{\rho(\tau-1,{\mathcal{B}})}$, since 
$\frac{\mu(S \cap
S_{\tau})}{\mu({\mathcal{B}})}=\rho(\tau,{\mathcal{B}})$ and
$\frac{\mu(S)}{\mu({\mathcal{B}})}=\rho(\tau-1,{\mathcal{B}})$.

The remaining calculations to arrive in Eq. (\ref{HKS_cond}) consider
the measure of the region $S_{\tau} \cap S_{\tau+P}$ (instead of $S
\cap S_{\tau}$) in order to have a positive condition probability,
i.e. $\frac{\mu(S_{\tau} \cap S_{\tau+P})}{\mu(S_{\tau})}>0$, with
$\mu(S_{\tau})$ representing the measure of the trajectories that
return to ${\mathcal{B}}$ after $\tau$ iterations and $\mu(S_{\tau}
\cap S_{\tau+P})$ the measure of the trajectories that return to
${\mathcal{B}}$ after $\tau + P$ iterations. The inequality in
Eq. (\ref{HKS_cond}) comes from the fact that one neglects the
infinitely many terms coming from the measure
$\mu_{REC}(\tau,{\mathcal{B}})$ that would contribute positively to
this summation.

\section{Estimation of errors in $H_{KS}$ and $\langle H_{KS} \rangle$} 

In order to derive Eq. (\ref{demonstration}), we have assumed that
$\int -\mu e^{-\mu \tau} log{[\mu e^{-\mu \tau}]} d\tau=
-\log{[\mu]}+1$, which is only true when $\tau_{min}$=0.  In reality,
for $\tau_{min}>0$, we have $\int_{\tau_{min}}^{\infty} -\mu e^{-\mu
  \tau} log{[\mu e^{-\mu \tau}]} d\tau$ = $e^{-\mu \tau_{min}}[\mu
\tau_{min} - \log{\mu}]+1$, but as $\epsilon$ tends to zero $\mu
\tau_{min} \rightarrow 0$ and therefore, as assumed $\int -\mu e^{-\mu
  \tau} log{[\mu e^{-\mu \tau}]} d\tau \approxeq -\log{[\mu]}+1$.

Making the same assumptions as before that $\rho(\tau,\epsilon)
\rightarrow \mu e^{-\mu \tau}$ as $\epsilon \rightarrow 0$, and using
Eq. (\ref{benoit}), then Eq.  (\ref{HKS_prob}) can be written as
\begin{equation}
  H_{KS}({\mathcal{B}}[\epsilon]) \approxeq \lambda + \frac{1}{\tau_{min}({\mathcal{B}}[\epsilon])}.
\label{erro}
\end{equation}

Theoretically, one can always imagine a region $\epsilon$ with an
arbitrarily small size, which would then make the term
$\frac{1}{\tau_{min}}$ to approach zero. But, in practice, for the
considered values of $\epsilon$, we might have (for atypical
intervals) shortest returns as low as $\tau_{min}=4$. As a result, we
expect that numerical calculations of the quantity in Eq.
(\ref{HKS_prob}) would lead us to a value larger than the positive
Lyapunov exponent, as estimated from the returns of the trajectory to
a particular region.

Naturally, $\frac{1}{\tau_{min}}$ would provide a local deviation of
the quantity in Eq. (\ref{HKS_prob}) with respect to the KS entropy.
To have a global estimation of the error we are making by estimating
the KS entropy, we should consider the error in the average quantity
$\langle H_{KS} \rangle$ which is given by
\begin{equation}
E=\sum_{\mathcal{B}(\epsilon)} \frac{1}{\tau_{min}({\mathcal{B}}[\epsilon])}
\label{erro_medio}
\end{equation}
\noindent
where the average is taken over $L$ different regions in phase space, and 
thus for chaotic systems with no more than one positive
Lyapunov exponent
\begin{equation}
\langle H_{KS} \rangle \approxeq \lambda + E
\label{erro_medio1}
\end{equation}

To generalize this result to higher dimensional systems, we make the
same assumptions as the ones to arrive to Eq. (\ref{erro}), but now we
use Eq. (\ref{demonstration}). We arrive that 
\begin{equation}
  \langle H_{KS}({\mathcal{B}}[\epsilon]) \rangle \approxeq H + E, 
\label{erro1}
\end{equation}
\noindent
where $H$ denotes the exact value of the KS entropy. 

Finally, it is clear from Eq. (\ref{erro1}) that $\langle
H_{KS}({\mathcal{B}}[\epsilon]) \rangle$ is an upper bound for the KS
entropy. Thus,
\begin{equation}
  H \leq \langle H_{KS}({\mathcal{B}}[\epsilon]) \rangle. 
\label{erro2}
\end{equation}

\section{Estimating the KS entropy and a lower bound of it in maps}

\begin{figure}
\centerline{\hbox{\psfig{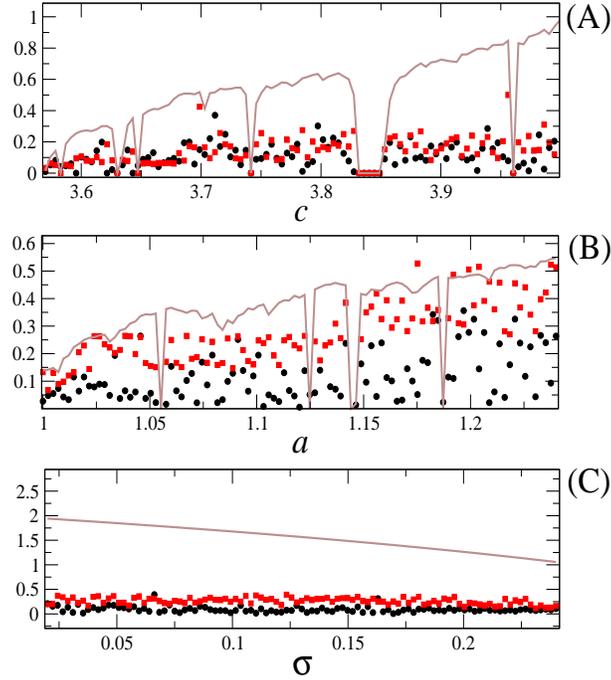}}}
\caption{[color online] Results from Eq. (\ref{HKS_cond}). 
The probability function $\rho(\tau,{\mathcal{B}})$ of the FPRs (RPs)
were obtained from a series of 500.000 FPRs (from a trajectory of
length 15.000 points).  The brown line represents the
values of the positive Lyapunov exponent. In (A) we show results for the Logistic map as
we vary the parameter $c$, $\epsilon_2=0.002$ for the black circles
and $\epsilon_1$=0.001 for the red squares. In (B) we show results for
the H\'enon map as we vary the parameter $a$ for $b$=0.3,
$\epsilon_2=[0.002-0.03]$ for the black circles and $\epsilon_1$=0.002
for the red squares, and in (C) results for the coupled maps as we
vary the coupling strength $\sigma$, $\epsilon_2$=0.05 for the black
circles and $\epsilon_1$=0.02 the red squares.}
\label{fig2}
\end{figure}

\begin{figure}[!h]
  \centerline{\hbox{\psfig{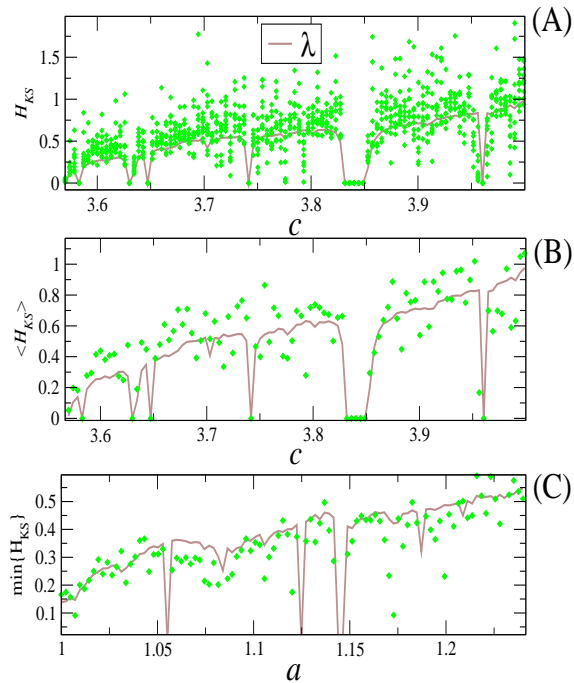}}}
  \caption{[color online] Results from Eq. (\ref{HKS_prob}) applied to
    the FPRs coming from the Logistic map (A-B), as we vary the
    parameter $c$ and $\epsilon_1$=0.00005, and from the H\'enon map
    (C), as we vary the parameter $a$ and $\epsilon_1=0.001$. These
    quantities were estimated considering 10 randonmly selected
    regions. The brown line represents the values of the positive
    Lyapunov exponent. The probability density function
    $\rho(\tau,{\mathcal{B}})$ was obtained from a series of 500.000
    FPRs. Green diamonds represent in (A) the values of $H_{KS}$
    calculated for each one of the 10 randonmly selected regions, in
    (B) the average value $\langle H_{KS} \rangle$ and in (C) the
    minimal value of $H_{KS}$.}
\label{fig3}
\end{figure}

In order to illustrate the performance of our formulas we use the
Logistic map [$x_{n+1}=cx_n(1-x_n)$], the H\'enon map
[$x_{n+1}=a-x_n^2+by_n$, and $y_{n+1}=x_n$], and a system of two
mutually coupled linear maps [$x_{n+1}=2x_n - 2 \sigma (y_n-x_n)$ and
$y_{n+1}=2y_n - 2 \sigma (x_n-y_n)$, $mod(1)$], systems for which
Pesin's equality holds.  The parameter $\sigma$ in the coupled maps
represents the coupling strength between them, chosen to produce a
trajectory with two positive Lyapunov exponents.

Using Eqs. (\ref{HKS_prob}) and (\ref{benoit}) to estimate $H_{KS}$
and $\lambda$ furnishes good values if the region ${\mathcal{B}}$
where the returns are being measured is not only sufficiently small
but also well located such that $\tau_{min}$ is sufficiently large. In
such a case the trajectories that produce such a short return visit
the whole chaotic set \cite{dimensions}. For that reason we measure
the FPRs for 50 different regions with a sufficiently small volume
dimension, denoted by $\epsilon_1$, and use the FPRs that produce the
largest $\tau_{min}$, minimizing $H_{KS}$. Since the lower bound of
$H_{KS}$ in Eq. (\ref{HKS_cond}) is a minimal bound for the KS
entropy, the region chosen to calculate it is the one for which the
lower bound is maximal. This procedure makes $H_{KS}$ and its lower
bound (calculated using the FPRs) not to depend on ${\mathcal{B}}$.

As pointed out in Ref. \cite{letellier}, one should consider volume
dimensions (also known as thresholds) which depend linearly on the
size of the attractor
\cite{dimensions}, in order to calculate the
Shannon's entropy.  In this work, except for the H\'enon map, we could
calculate well $H_{KS}$, $\lambda$ and a lower bound for $H_{KS}$ from
the FPRs and RPs, considering for every system fixed values
$\epsilon_1$ and $\epsilon_2$. For the H\'enon map, as we increase the
parameter $b$ producing more chaotic attractors, we increase linearly
the size of the volume dimension $\epsilon_2$ within the interval
$[0.002-0.03]$.

We first compare $H_{KS}$ (see Fig. \ref{fig1}), calculated from
Eq. (\ref{HKS_prob}) in terms of the probabilities coming from the
FPRs and RPs, in green diamonds and brown stars, respectively, with
the value of the KS entropy calculated from the sum of the positive
Lyapunov exponents, represented by the brown straight line.  As
expected $H_{KS}$ is close to the sum of all the positive Lyapunov
exponents. When the attractor is a stable periodic orbit we obtain
that $H_{KS}$ is small if calculated from the RPs. In such a case, we
assume that $H_{KS}=0$ if calculated from the FPRs. This assumption
has theoretical grounds, since if the region is centered in a stable
periodic attractor and $\epsilon_1
\rightarrow 0$ (what can be conceptually make), one will clearly
obtain that the attractor is periodic.

\begin{figure}
  \centerline{\hbox{\psfig{file=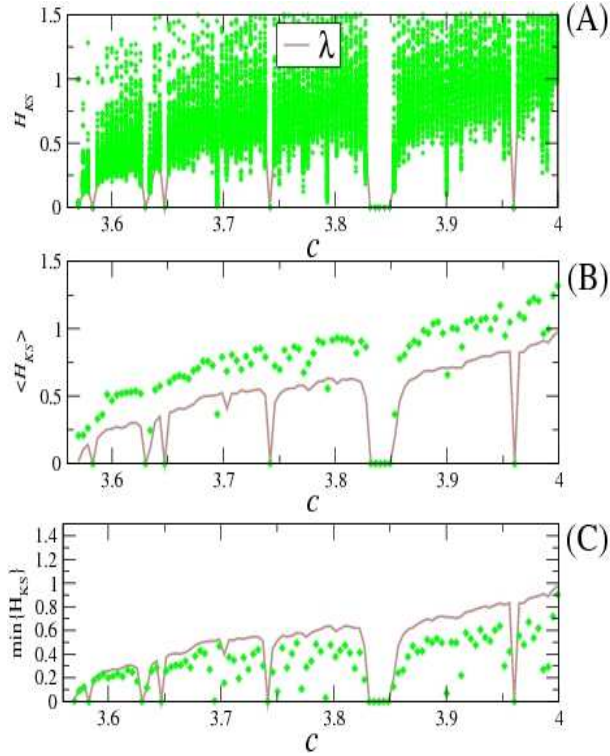,height=10.0cm,width=8.0cm}}}
\caption{[color online] The same quantities shown in Fig. \ref{fig3}, but now 
considering only the Logistc map, with $\epsilon_1$=0.0002 and 500
randonmly selected regions.}
\label{fig4}
\end{figure}

The value of the Lyapunov exponent calculated from the formula
(\ref{benoit}) is represented in Fig. \ref{fig1} by the blue up
triangles. As it can be checked in this figure, Eq. (\ref{benoit})
holds only for 1D hyperbolic maps. So, it works quite well for the
logistic map (a 1D ``almost'' uniformly hyperbolic map) and somehow
good for the H\'enon map. However, it is not appropriate to estimate
the sum of the positive Lyapunov exponents coming from 2D coupled
systems. This formula assumes sufficient hyperbolicity and
one-dimensionality such that $e^{\tau_{min} \lambda}=1/\epsilon$.

To compare our approach with the method in Ref. \cite{grassberger},
we consider the H\'enon map with $a$=1.4 and $b=0.3$ for which the
positive Lyapunov exponent equals 0.420. Therefore, by using Ruelle
equality, $H_{KS}=0.420$.  In Ref. \cite{grassberger} it is obtained
that the correlation entropy $K_2$ equals 0.325, with $H_{KS} \geq
K_2$ and in Ref. \cite{cohen} $H_{KS}=0.423$. From
Eq. (\ref{HKS_prob}), we obtain $H_{KS}=0.402$ and from
Eq. (\ref{HKS_cond}), we obtain $H_{KS}
\geq 0.342$, for $\epsilon_1$=0.01.

In Fig. \ref{fig2}(A-C), we show the lower bound estimation of
$H_{KS}$ [in Eq. (\ref{HKS_cond})] in terms of the RPs (black circles)
and in terms of FPRs (red squares). As expected, both estimations
follow the tendency of $H_{KS}$ as we increase $a$.

Another possible way Eq. (\ref{HKS_prob}) can be used to estimate the
value of the KS-entropy is by averaging all the values obtained for
different intervals, the quantity  $\langle H_{KS} 
\rangle$ in Eq. (\ref{average_HKS}). In Fig. \ref{fig3}(A), we show the values 
of $H_{KS}$ as calculated from Eq. (\ref{HKS_prob}) considering a
series of FPRs with 500.000 returns of trajectories from the Logistic
map.  For each value of the control parameter $c$, we randomnly pick
10 different intervals with $\epsilon_1$=0.00005.  The average
$\langle H_{KS} \rangle$ is shown in Fig. \ref{fig3}(B). As one can
see, $\langle H_{KS} \rangle$ is close to the Lyapunov exponent
$\lambda$.  Notice that from Fig. \ref{fig3}(A) one can 
see that the minimal value of $H_{KS}$ (obtained for the largest
$\tau_{min}$) approaches well the value of $\lambda$.

In order to have a more accurate estimation of the KS-entropy for the
H\'enon map, we have used in Figs.  \ref{fig1}(B) and \ref{fig2}(B) a
varying $\epsilon_2$ depending on the value of the parameter $a$,
exactly as suggested in
\cite{letellier}, but similar results would be obtained considering a
constant value. As an example, in Fig. \ref{fig3}(C) we show the minimal
value of $H_{KS}$ considering regions with $\epsilon_1=0.001$, for a 
large range of the control parameter $a$.  

In order to illustrate how the number of regions as well
as the size of the regions alter the estimation of the KS-entropy, we
show, in Fig.
\ref{fig4}(A-C), the same quantities shown in Fig. \ref{fig3}(A-B), 
but now from FPRs exclusively coming from the Logistic map,
considering 500 randonmly selected regions all having sizes 
$\epsilon_1$=0.0002. Recall that in Figs. \ref{fig1} and \ref{fig3},
the minimal value of $H_{KS}$ was chosen out of no more than 50
randonmly selected regions. Comparing Figs.  \ref{fig3}(B) and
\ref{fig4}(B) one notices that
an increase in the number of selected regions is responsible to smooth
the curve of $\langle H_{KS} \rangle$ with respect to $c$. Concerning
the minimal value of $H_{KS}$, the use of intervals with size
$\epsilon_1=0.0002$ provides values close to the Lyapunov exponent if
this exponent is sufficiently low (what happens for
$b<3.7$). Otherwise, these values deviate when this exponent is larger
(what happens for $b > 3.7$). This deviation happens because for these
chaotic attractors the size of the chosen interval was not
sufficiently small \cite{dimensions}.

Notice that the estimated KS entropy deviates from $\lambda$.  See,
for example, Figs.  \ref{fig3}(B) and \ref{fig4}(B).  One sees two
main features in these figures. The first is that for most of the
simulations, $\langle H_{KS} \rangle > \lambda$.  The second is that
the larger $\lambda$ is, the larger the deviation is.  The reason for
the first feature can be explained by Eqs. (\ref{erro_medio1}) and
(\ref{erro2}). The reason for the second is a consequence of the fact
that the larger the Lyapunov exponent is, the smaller $\tau_{min}$ is,
and therefore the larger the error in the estimation of the KS
entropy.

\begin{figure}
  \centerline{\hbox{\psfig{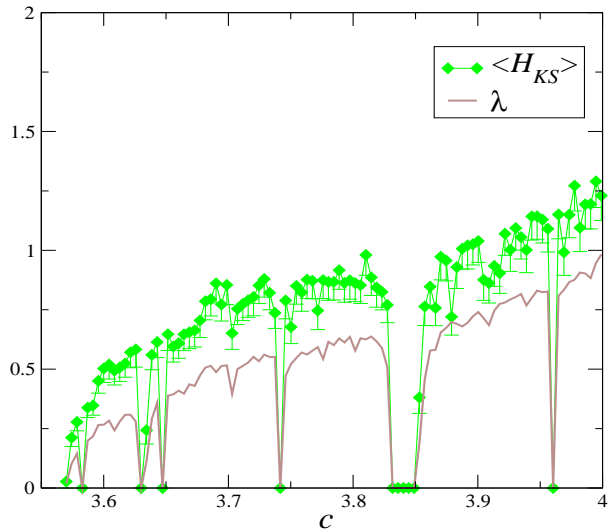}}}
  \caption{[color online] Results obtained considering FPRs coming
    from the Logistic map, as we vary the parameter $c$ and
    $\epsilon_1$=0.0002. The probability density function
    $\rho(\tau,{\mathcal{B}})$ was obtained from a series of 250.000
    FPRs. Green diamonds represent the values of $\langle H_{KS}
    \rangle$ calculated for each one of the 100 randomly selected
    regions. The error bar indicates the value of the average error
    $E$ in Eq. (\ref{erro_medio1}). These quantities were estimated
    considering 100 randomly selected regions. The brown line
    represents the values of the positive Lyapunov exponent. }
\label{fig5}
\end{figure}

To see that our error estimate provides reasonable results, we
calculate the quantities $\langle H_{KS} \rangle$ (green diamonds in
Fig. \ref{fig5}), for the Logistic map considering a series of 250.000
FPRs to $L$=100 randomly selected regions of size $\epsilon_1=0.0002$,
and the average error $E$, in Eq.  (\ref{erro_medio1}) [shown in Fig.
\ref{fig5} by the error bars].  The value of the positive Lyapunov
exponent is shown in the full brown line.

The error in our estimation is inversely proportional to the shortest
return. Had we considered smaller $\epsilon$ regions, $\tau_{min}$
would be typically larger and as a consequence we would obtain a
smaller error $E$ in our estimation for the KS entropy. Had we
consider a larger number of FPRs, the numerically obtained value of
$\tau_{min}$ would be typically slightly smaller, making the error $E$
to become slightly larger. So, the reason of why the positive Lyapunov
exponent in Fig.  \ref{fig5} is located bellow the error bars for the
quantity $\langle H_{KS} \rangle$ is a consequence of the fact that we
have only observed 250.000 returns, producing an overestimation for
the value of $\tau_{min}$. Had we considered a larger number of FPRs
would make the error $E$ to become slightly larger.

The considered maps are Ergodic. And therefore, the more (less)
intervals used, the shorter (the longer) the time series needed in
order to calculate the averages from the FPR as well as from the RP,
as the average $\langle H_{KS} \rangle$.

\section{Conclusions}

Concluding, we have shown how to estimate the Kolmogorov-Sinai entropy
and a lower bound of it using the Poincar\'e First Return Times (FPRs)
and the Recurrence Plots. This work considers return times in discrete
systems. The extension of our ideas to systems with a continuous
description can be straightforwardly made using the ideas in Ref.
\cite{gao}.

We have calculated the expected error in our estimation for the KS
entropy and shown that this error appears due to the fact that FPRs
can only be physically measured considering finite sized regions and
only a finite number of FPRs can be measured. This error is not caused
by any fundamental problems in the proposed Eq.  (\ref{HKS_prob}).
Nevertheless, even for when such physical limitations are present, the
global estimator of the KS entropy [Eq.  (\ref{average_HKS})] can be
considered as an upper bound for the KS entropy [see Eq.
(\ref{erro2})].


\begin{thebibliography}{99}
\bibitem{KS-entropy}  The time version of the KS entropy is
calculated from the average value of the difference between the
Shannon entropies (per unit of time) of a trajectory with a length $n
\delta t$ and of a trajectory with a length $(n+1) \delta t$, for all
possible values of $n$. More rigorously, $\lim_{n \rightarrow \infty}
\lim_{(\epsilon,\delta t)
\rightarrow 0} \frac{1}{n \delta \tau} \sum_{i=0}^{n-1} [K_{i+1}-K_{i}]$, 
where $K_{i}$ is the shannon's entropy of a trajectory that visits
$(n+1)$ volumes of sides $\epsilon$ in phase space. In each
volume the trajectory remains during a time interval $\delta t$. As $i
\rightarrow \infty$, the terms $K_{i+1}$ are infinities, but not the
difference $[K_{i+1}-K_{i}]$. Since $\sum_{i=0}^{n-1} [K_{i+1}-K_{i}]
= K_{n} - K_{0}$ and that $K_{0}/{n
\delta \tau} \rightarrow 0$, then 
$H_{KS} = \lim_{n \rightarrow \infty}
\lim_{(\epsilon,\delta t)
\rightarrow 0} \frac{1}{n \delta \tau}  K_{n}$. So, basically the 
time version of the KS-entropy can be thought as the Shannon's
entropy divided by a characteristic time, $T$, yet to be determined.
\bibitem{coment_equivalencia} 
While the white vertical segments correspond to the first Poincar\'e
returns to an interval centered at a point, the white diagonal
segments provide the $n$-th Poincar\'e returns. As an example, imagine
that there are two black points placed at the coordinates (10,20) and
(20,30) in the RP. In the white diagonal segment connecting these two
black points there are no black points, which means that we have a
nonrecurrent diagonal segment of length $10$.  That can only be
possible if two first Poincar\'e returns of length 10 happened, or if
one second Poincar\'e return of length 10 happened.
\bibitem{shannon_entropy_FPR} 
Assuming that $\rho(\tau,{\mathcal{B}}) = \mu e^{-\mu \tau}$, this leads 
$H_S = -\log{(\mu)}+1$. Since $\mu \sim \epsilon$, if
$\epsilon \rightarrow 0$, then $H_S \rightarrow \infty$.
\bibitem{dimensions} Usually, the larger an attractor, the larger the
  Lyapunov exponent. As discussed in \cite{bruin}, $1=\epsilon
  e^{\tau_{min} \lambda}$. Assuming that a sufficiently small
  threshold $\epsilon$ provides a sufficiently large $\tau_{min}$,
  this leads to $\epsilon = e^{-\tau_{min} \lambda}$ and therefore the
  more chaotic an attractor, the larger $\epsilon$ must be in order to
  have a sufficiently large $\tau_{min}$. Since the series of FPRs can
  be (in principle theoretically) calculated from arbitrarily large
  trajectories, we can consider (from a theoretical
  perspective) regions sufficiently small so that one can obtain FPRs
  with sufficiently large $\tau_{min}$, even for different values of
  the control parameters.  On the other hand, RPs are constructed with
  trajectories (time series) shorter than the ones used for the FPRs.
  In order to have a RP for which $\langle \tau_{min} \rangle$ is
  sufficiently large and at the same time producing a reasonable
  continuous distribution $\rho(\tau)$, the volume dimensions
  $\epsilon_2$ considered to construct the RPs should be reasonably
  larger than $\epsilon_1$.  In addition, as one changes the control
  parameters producing more complex chaotic attractors, we might have
  to increase the value of $\epsilon_2$.
\end{thebibliography}
\end{document}